\documentstyle[12pt]{article}

\newcommand{\del}{\delta}
\newcommand{\eps}{\varepsilon}

\newcommand{\beq}{\begin{equation}}
\newcommand{\eeq}{\end{equation}}
\newcommand{\ba}{\begin{array}}
\newcommand{\ea}{\end{array}}
\newcommand{\beqa}{\begin{eqnarray}}
\newcommand{\eeqa}{\end{eqnarray}}
\newcommand{\bd}[1]{ \mbox{\boldmath $#1$}  }
\newcommand{\ti}{\tilde}
\begin{document}

\begin{center}
{\bf\Large Dual Killing-Yano symmetry and multipole moments in
electromagnetism and mechanics of continua}
\end{center}
\begin{center}
{\bf D. B\u aleanu}\footnote[1]{ Permanent address:Institute of
Space Sciences,
 P.O.BOX, MG-23, R 76900, Magurele-Bucharest, Romania,
} \footnote[2]{E-mail:
 baleanu@venus.nipne.ro,dumitru@newton.physics.metu.edu.tr}\\
 {\em Middle East Technical
University,\\
Physics Department
 -06531 Ankara,Turkey}\\ {and}\\
{\em Bogoliubov Laboratory of Theoretical Physics,\\
 141980 Dubna,Moscow Region,Russia}\\
 {\bf and}\\
  {\bf V.M. Dubovik, \c S. Mi\c sicu}\footnote[3]{
Permanent address:Department for Theoretical Physics, National
Institute for Nuclear Physics, POB MG-6, Bucharest-M\u agurele,
Romania}\\

     {\em Bogoliubov Laboratory of Theoretical Physics,\\
 141980 Dubna,Moscow Region,Russia}\\
\end{center}

\begin{abstract}
In this work we introduce the Killing-Yano symmetry on the phase
space and  we investigate the symplectic structure on the space of
Killing-Yano tensors. We  perform the detailed analyze of the
$n$-dimensional flat space and the Riemaniann manifolds with constant 
scalar  curvature. We investigate the form of some multipole
tensors, which arise in the expansion
of a system of charges and currents, in terms of second-order
Killing-Yano tensors in the phase space of classical mechanics.
 We find  some relations between  these tensors and the generators
of dynamical symmetries like the angular momentum, the
mass-inertia tensor, the conformal operator and the momentum
conjugate Runge-Lenz vector.

\end{abstract}

\newpage

\section{Introduction}

 Killing tensors are indispensable tools in the quest for exact
 solutions in many branches of general relativity as well as classical
mechanics.
 Killing tensors can also be important for solving the equations of
motion in particular space-time \cite{kerr}. Killing-Yano tensors
were introduced in 1952 by  Yano
 from a mathematical point of view on the configuration space \cite{yano}.
 When a manifold admits a Killing-Yano
tensor we can construct a Killing tensor and a new constant of
motion in the case of geodesic motions \cite{hol,dani}. It was a
big success of Gibbons  et all. to have been able to show that
Killing-Yano tensors, which had long been known for relativistic
systems as a rather mysterious structure, can be understood as an
object generating a ''non-generic symmetry'', i.e. a supersymmetry
appearing only in the specific space-time \cite{Gi}. On the other
hand Lax pairs tensors can be viewed as a generalization  of
Killing-Yano tensors \cite{Lax}. The relation
 between Killing-Yano tensors and Nambu tensors was found in \cite{bal}.
  As it is well known in the expansion of charge and currents
 in electromagnetism,
three families of multipole moments arise : the charge, the
magnetic and the toroid moments \cite{DC74}. Among the first
members of these multipolar families, the time derivative of the
charge dipole $\dot{\bd{d}}$, charge quadrupole $\dot{Q}_{ij}$ and
the magnetic dipole $\bd{\mu}$, correspond to infinitesimal
translations, shears and rotations of the points of a continuous
distribution of charged matter. For example the charge multipole
moments, $Q_{i_1i_2\cdots i_n}$, are related to the $n$-th order
inertia moments of a continuous distribution of mass
\cite{Chan68}. In view of the correspondence between the electric
charge $e$, which is connected to gauge invariance and the
gravitational mass $m$, which is related to the Poincar\' e
invariance, we make the formal change of the current density
$\bd{j}$ by the momentum vector \bd{p}. In this way we obtain the
following associations for these tensors \beqa
 \dot{d_i} &  \longrightarrow &  p_i, \\ {Q_{ij}}&
\longrightarrow & x_i x_j-{1\over3}\bd{r}^2\delta_{ij},\\
\dot{Q_{ij}}& \longrightarrow & x_ip_j+x_jp_i-{2\over3}
(\bd{r}\cdot\bd{p})\delta_{ij},\\ \mu_i &  \longrightarrow & L_i.
\eeqa
 These tensors can be found as generators of many Lie
dynamical symmetries like for example the three-dimensional
rotation group $SO$(3), which is
 generated by the three
components of the angular momentum $L_i$, the group of the rigid
rotator $Rot$(3), generated by the mass quadrupole tensor $Q_{ij}$
and $L_i$ \cite{Ui70}, or the linear motion group $SL$(3) which in
turn is generated by the shear tensor $S_{ij}\equiv\dot{Q}_{ij}$
and $L_i$ \cite{cus68}. It is then natural to seek for
 the symmetries and the geometrical features of the
higher-rank tensors arising in the multipole expansion. An
important point that one should mention is that the above tensors
are written in the configuration space.
 Unfortunately the components of
the higher-rank multipoles do not satisfy the closure relations
for the Lie symmetry. For our purposes it will turn out to be
useful to consider the same tensors in the momentum space too. It
is more convenient to write these tensors in terms of purely
geometric quantities in  a form which will allow the
generalization to higher dimensions. On the other hand the
Killing-Yano symmetry was defined only on the configuration space.

For all these reasons the extension of the Killing-Yano symmetry
on the phase space is very interesting to investigate.

 The plan of the paper is as follows:

 In Section 2 the extension of the  Killing-Yano symmetry on the phase space is presented.
 In Section 3 multipole and dynamical
 symmetry tensors are investigated using dual Killing-Yano tensors.
 In Section 4 we present our concluding remarks.

\section{Dual Killing-Yano symmetry}
A Killing-Yano tensor \cite{yano} is an antisymmetric tensor which
satisfies\\
 the equation  \beq  D_{\lambda}f_{\nu\mu} +
D_{\nu}f_{\lambda\mu}=0 \label{dan1}. \eeq Here  $D$ denotes the
covariant derivative.

For a given metric $g_{\mu\nu}$ instead of $x_{\mu}$ we consider
the momentum $p_{\mu}$. In this way we have obtained the metric
${\tilde g_{\mu\nu}}$ on the momentum space.
 Performing the operation of mapping $x_{\mu}$ to $p_{\mu}$ twice,
leads back to the original metric $g_{\mu\nu}$.
 We call ${\tilde
g_{\mu\nu}(p)}$ dual to $g_{\mu\nu}(x)$.

 The existence of a Killing-Yano tensor on a given
manifold is deeply related to the existence of a new supersymmetry
in  the case of geodesic motion on the spinning space \cite{hol}. We
know that
 the action for the geodesic of spinning space has the form
 \cite{hol}
\begin{equation}\label{spin}
S=\int_a^bd\tau \left( \,{\frac 12}\,g_{\mu \nu }(x)\,\dot{x}^\mu \,\dot{x}%
^\nu \,+\,{\frac i2}\,g_{\mu \nu }(x)\,\psi ^\mu \,{\frac{D\psi
^\nu }{D\tau }}\right) .
\end{equation}
 The overdot denotes an ordinary proper-time
derivative $d/d\tau $ whilst the covariant derivative of a
Grassmann variable $\psi ^\mu $ is defined by
 $D\psi ^\mu/D\tau
=\dot{\psi}^\mu +\dot{x}^\lambda \,\Gamma _{\lambda \nu
}^\mu\,\psi ^\nu$ .
 In general, the symmetries of a spinning-particle model can
be divided into two classes.
 First, there are conserved quantities
which exist in any theory and these are called
 {\it  generic} constants of motion .
 It has been shown that for a spinning particle model defined by
the action ( \ref{spin}) there are four generic symmetries (for
more details see  \cite{Gi}).By construction we have obtained four
{\it generic} symmetries on the momentum space too. The second
kind of conserved quantities, called {\it non-generic}, depend on
the explicit form of the metric $g_{\mu \nu }(x)$.
 The existence of a Killing-Yano tensor $f_{\mu\nu}$ of the bosonic
manifold is equivalent to the existence of a supersymmetry for the
spinning particle with supercharge
$Q_f=f^{\mu}_{a}\Pi_{\mu}\psi^{a}-\frac{1}{3}iH_{abc}\psi^{a}\psi^{b}\psi^{c}$
satisfies $\{Q,Q_f\}=0$, where $H_{\mu\nu\lambda}=
D_{\lambda}f_{\mu\nu}$ , $\Pi_{\mu}=g_{\mu\nu}{\dot x^{\mu}}$
whereas the supercharge $Q$ has the form
 $Q=\Pi_{\mu}\psi^{\mu}$ (see
\cite{Gi} for more details). Because the dual metric $\tilde
g_{\mu\nu}$ admits a Killing-Yano tensor $\tilde f_{\mu\nu}$ the
corresponding {\it non-generic} supersymmetries  is defined by the
supercharge ${\tilde Q_f=f^{\mu}_{a}}{\tilde
\Pi_{\mu}}\psi^{a}-\frac{1}{3}i{\tilde
H_{abc}}\psi^{a}\psi^{b}\psi^{c}$ satisfying
 $\{\tilde Q,\tilde Q_f\}=0$.
 Here
 ${\tilde
H_{\mu\nu\lambda}}= {\tilde D_{\lambda}}{\tilde f_{\mu\nu}}$, the
canonical momentum is   ${\tilde \Pi_{\mu}}={\tilde
g_{\mu\nu}}{\dot p^{\mu}}$ and the supercharge $Q$ has the form
$\tilde Q=\tilde \Pi_{\mu}\psi^{\mu}$ .
  We mention that ${\tilde D_{\lambda}}$ means the covariant
derivative on the momentum space.

 Using dual Killing-Yano symmetry we have obtained a pair of Killing-Yano
tensors $(f,\ti{f})$ defined on the phase space.
\subsection{\bf {Examples}}

{\bf 2.1.1.Flat space case}\\

 In the case of
three-dimensional flat space these tensors have the following form
: \beq f_{ij}=\eps_{kij}x_k\quad \tilde{f_{ij}}=\eps_{kij}p_k
\label{dan2}. \eeq Eq.({\ref{dan2}}) can be reversed and thus one
may express the position $x$ and momentum $p$ variables in terms
of  $f$ and $\tilde{f}$ \beq
 x_{i}={1\over 2}\eps_{ijk}f_{jk},\qquad
p_{i}={1\over 2}\eps_{ijk}{\tilde f}_{jk} \label{7}. \eeq

 The Poisson bracket of  $f$ and $\tilde{f}$ reads
\beq \{f_{ij}, {\tilde f}_{kl} \} = \del_{ik}\del_{jl}-
\del_{il}\del_{jk}. \eeq
 A scalar product can be defined for these Killing-Yano tensors.
    For example, the square of $f$ can be written as follows
\beq f^2\equiv f\cdot f\equiv f_{ij}f_{ij}. \eeq Equation
(\ref{7}) enables us to construct the phase-space in terms of the
Nambu tensor $\epsilon_{ijk}$ and the Killing-Yano tensors
$f_{ij}$ and ${\ti f}_{ij}$. When a manifold admits a Killing-Yano
tensor $f_{ij}$ then we can construct a Killing tensor
$K_{ij}=x_{i}x_{j}-r^{2}\delta_{ij}$. This Killing tensor
corresponds to a constant of motion $K=K_{ij}p_{i}p_{j}$. The 
Nambu tensor $\epsilon_{ijk}$ is of rank three and it
defines a Nambu mechanics with the constants of motion $H=p^2$
and $K$ \cite{nambu,takh}.
 These results can be  generalized in the flat space of an arbitrary
dimension \cite{bal}. In this case we have
 \beq
x_{i}={1\over n!}\epsilon_{i_{1}\cdots i_{n}} f_{i_{1}\cdots
i_{n}}\quad ,p_{i}={1\over n!}\epsilon_{i_{1}\cdots i_{n}} {\ti
f}_{i_{1}\cdots i_{n}} \label{10}. \eeq
 Equation (\ref{10}) enables
us to construct the phase space in terms of Nambu tensor
$\epsilon_{i_{1}\cdots i_{n}}$  and Killing-Yano tensors
$f_{i_{1}\cdots i_{n}}$ and $\tilde f_{i_{1}\cdots i_{n}}$.\\

{\bf 2.1.2.Riemannian manifold with constant scalar curvature}\\

 It is well known  that  any $n$-dimensional Riemannian
manifold with constant scalar curvature admits ${n(n-1)\over 2}$
Killing-Yano tensors of order two \cite{kram}.
   For example in the three dimensional case,
the corresponding metric with constant curvature
 has the form
\beq
 ds^{2}=\left(1 +{Kr^{2}\over
4}\right)^{-2}\sum_{i=1}^{3}\left(dq^{i}\right)^{2}. \eeq where
 $r=\sqrt{\sum_{i=1}^{3}\left(q^{i}\right)^2}$ and
 $q^{i}(i=1,2,3)$ are the coordinates, whereas
$K$ is a real constant denoting curvature of the configuration
space \cite{katayama}. In this case we have  three Killing-Yano
tensors $f_{\mu\nu}$ and three ${\tilde f_{\mu\nu}}$. When $q^i$
are the spherical coordinates the expressions of the Killing-Yano
tensors looks like
{\small
 \beqa
&f_{12}=&{{r\sin\phi\over 16(1+{Kr^2\over 4})^2}},
f_{13}={r\sin2\theta\cos\varphi\over 32(1+{Kr^2\over 4})^2},
f_{23}= {r^2\sin\theta^2\cos\varphi({Kr^2-4})\over (4 +
Kr^2)(1+{Kr^2\over 4})^2},\cr &\tilde f_{12}=&{16p\sin\varphi
\over (1+{Kp^2\over 4})^2}, \tilde
f_{13}={p\sin2\theta\cos\varphi\over 32(1+{Kp^2\over 4})^2} ,
\tilde f_{23}={p^2\sin\theta^2\cos\varphi(Kp^2-4)\over
(1+{Kp^2\over 4})^2(4+Kp^2)}. \eeqa
}
  We  are able to express the components of
Runge-Lenz vector and the energy level of the Kepler problem
\cite{katayama} in terms of purely geometric quantities
$f_{\mu\nu}$ and ${\tilde f_{\mu\nu}}$.

\subsection{\bf {Symplectic structure}}

In this subsection the symplectic structure to the space of
Killing-Yano tensors is constructed.
 Let us consider for the beginning
 the $n$-dimensional flat space case. From (\ref{dan1}) we found
 that the
 Killing-Yano tensors $f$ and $\tilde f$ are
$
f_{i_{1}\cdots i_{n-1}}=\epsilon_{i_{n}\cdots i_{1}\cdots
i_{n-1}}x^{i_{n}}$,
$
{\tilde f_{i_{1}\cdots i_{n-1}}}=\epsilon_{i_{n}\cdots i_{1}\cdots
i_{n-1}}p^{i_{n}}$.
  Because each of the antisymmetric tensors $f_{i_{1}\cdots i_{n-1}}$ and
$\tilde f_{i_{1}\cdots i_{n-1}}$ has  $n$ independent components we
can consider $\bd{f}$  and $\bd{\tilde f}$ as a n-dimensional
vectors.  We can combine the vectors $\bf f$ and ${\bf {\tilde
f}}$ into a 2$n$ dimensional vector ${\bf x}=(\bd{f},\bd{\tilde f})$,
interpret the quantities 
$({\partial H\over\partial\bd{\tilde f_{j}}},
  {\partial H\over\partial\bd{f_{k}}})$, as a 2$n$-dimensional
vector $\bd{\nabla} H$, and introduce a $2n\times 2n$ matrix 
$ J=\pmatrix{0 &-I \cr
             I &0 \cr}$,
where I is the $n\times n$ identity matrix. With this notation
Hamilton's equations can be unified in the form
  ${\dot{\bf x}}=J.\bd{\nabla} H(\bd{x})$.

   Let us consider a even-dimensional Riemannian manifold  having
constant scalar curvature. It is well known that all symplectic
structures have locally the same structure.

A precise formulation of this assertion is given by Darboux's
theorem \cite{darb}. Due to this theorem, any statement of local
nature which is invariant under symplectic transformation and has
been proved for the standard phase space $(M={\bd
R}^{2n},\omega=\sum_{k=1}^{n}dp_{j}\wedge dq^j)$ can be extended
to all symplectic manifolds.
 On the other hand from equation (\ref{dan1}) we
found that an antisymmetric covariant constant tensor $f_{\mu\nu}$
is a solution of Killing-Yano equations. If the corresponding form
is non-degenerate then we have a symplectic structure on this
manifold. By construction the dual manifold admits a symplectic
structure. If a manifold admits a non-degenerate covariant
constant Killing-Yano, then we can construct a symplectic structure
to the space of Killing-Yano tensors.
 As an example we
mention here the self-dual Taub-NUT metric \cite{bali}. In this
case we have four Killing-Yano tensors. Three of these, denoted by
$f_i$ are special because they are covariant constant and
non-degenerate. The dual manifold  has three covariant constant
non-degenerate Killing-Yano tensors $\tilde f_{i}$.  In the
two-form notation, using the spherical co-ordinates $(r,\theta
,\varphi )$ and respectively $(p,\theta,\varphi)$ the explicit
expressions are: \beqa f_i&=&4m(d\psi +\cos \theta d\varphi
)dx_i-\epsilon _{ijk}(1+{\frac{2m}r} )dx_j\wedge dx_k ,\cr
{\tilde f_i}&=&4m(d\psi +\cos \theta d\varphi )dp_i-\epsilon _{ijk}(1+{\frac{2m}p}%
)dp_j\wedge dp_k . \eeqa

\section{Multipole and dynamical symmetry tensors}

 In this section  the expressions of the multipole tensors in
 terms of purely geometric quantities $(f,\tilde f)$ are presented.
  The first step consists in writing  the square of the radius
$\bd{r}^2$ and of the impulse $\bd{p}^2$ in terms of $(f,\tilde
f)$. \beq \bd{r}^2={1\over 2}f^2\quad \bd{p}^2={1\over 2}{\ti
f}^2, \eeq

The magnetic dipole tensor is given by \beq \mu_i = L_{i} =
{1\over 2}\eps_{klm}f_{ki}{\tilde f}_{lm}, \eeq and the dilatation
has the form \beq D \equiv \bd{r}\cdot\bd{p}= {1\over
2}f_{ij}{\tilde f}_{ij}. \eeq The quadrupole mass-inertia tensor
reads \beq Q_{ij}={1\over 4}( f_{im}f_{mj} - {1\over
3}\del_{ij}f^2 ). \eeq The toroid dipole tensor, a quantity
related to the poloidal currents on a torus, can be written in the
following manner \cite{DTOS83} \beq T_{i}={1\over 10} ( x_i D -
2\bd{r}^2p_i ) = {1\over 40} \eps_{ijk}\{f_{jk}(f\cdot{\tilde f})
- 2{\ti f}_{jk}f^2\}. \eeq
 In the case of purely transversal velocity fields this expression
gets a  simplified form \cite{DT90}: \beq
 T_{i} = {1\over 2}d_i
D = {1\over 8}\eps_{ijk}f_{jk}(f\cdot{\tilde f}). \label{tor} \eeq
  Next we pass to other tensors, related to dynamical symmetries.
Consider first the conformal operator
\beq
 C_{i}= 2x_i D -
\bd{r}^2p_i = {1\over 4}\eps_{ijk}\{2f_{jk}(f\cdot{\tilde f}) -
{\ti f}_{jk}f^2\}. \eeq
  Together with the angular momentum $L_i$, $C_i$ is a generator of a
symmetry group which obeys commutation relations isomorphic to
those of $SO$(4). This is a subgroup of the group $SO$(4,2)
\cite{Mur53,BR74} (isomorphic to the conformal group in Minkowski
space) which leaves invariant the free Maxwell's equations
\cite{Fush83}. The other generators of this larger group are the
impulse $p_i$ and the dilatation $D$ which were defined above.

Another interesting tensor is the following particular form of the
Runge-Lenz vector \cite{BR74} with components \beq
 A_i={1\over 2}x_ip^2 - p_iD-{1\over 2}x_i.
\label{runge}
\eeq
 This vector, together with the orbital angular
momentum $L_i$, the dilatation $D$ and other two vectors and two
scalars  generates the $SO$(4,2) group which contains as a subgroup
the symmetry group of the Hydrogen atom, i.e. $SO$(4). Thus, from the
algebraic point of view the properties of the Runge-Lenz vector
are similar to those of the conformal one. If next we take the
 {\em momentum conjugate} of (\ref{runge}), we  obtain
the following tensor in Killing-Yano form \beq {\tilde
A}_i={1\over 8}\eps_{ijk} \{(f^2-2){\tilde f}_{jk}-2f_{jk}(f\cdot
{\ti f}) \}. \eeq
  This tensor can be viewed as a symmetry generator like $A_i$,
but in the momentum space.
  Then we have obtained the following formula for the Killing-Yano
tensors in terms of the Runge-Lenz vectors and the conformal
operator in the space and momentum subspaces. \beq
f_{jk}=-\eps_{ijk}(2{\tilde A}_{i}+C_i)~~~~~~~, {\ti
f}_{jk}=-\eps_{ijk}(2A_{i}+{\ti C}_i). \label{fac} \eeq In this
way the toroid dipole tensor ({\ref{tor}}) can be directly related
to $SO$(4,2) symmetry generators in the  phase-space : \beq T_i =
(2{\ti A}_i+C_i)D. \eeq

When we move to the next rank multipolar tensors we encounter
the charge octupole tensor
\beq
Q_{ijk}={1\over 4}\left\{ \eps_{imn}(\del_{jk}f^2+2f_{jl}f_{lk}) -
{1\over 5}f^2(\eps_{imn}\del_{jk}+\eps_{jmn}\del_{ik}+\eps_{kmn}\del_{ij})
\right\},
\eeq
             the magnetic quadrupole tensor
\beq
\mu_{ij} = {1\over 3}(x_iL_j+x_jL_i) = -{1\over 3}
\left  \{
f_{ik}f_{kl} {\tilde f}_{lj}+f_{jk}f_{kl} {\tilde f}_{li}
\right \}  ,
\eeq
             and the toroid quadrupole tensor
\beq T_{ij}=\left ( f_{im}f_{mj} - {1\over 4}\delta_{ij}f^2 \right
)(f\cdot{\tilde f}) - {5\over 2}f_{im} {\tilde f}_{mj}f^2. \eeq
  Using again (\ref{fac}) we write the last tensor in terms of
dynamical symmetry generators, in the position subspace of
$Rot$(3) and $SO$(4,2), for the purely transversal gauge mentioned
above \beq T_{ij} = Q_{ij}D = {1\over 8}\left ( f_{im}f_{mj} -
{1\over 3}\delta_{ij}f^2 \right)(f\cdot{\tilde f}). \eeq

\section{Concluding remarks}

  In this paper the Killing-Yano symmetry was generalized to the
  phase space.
    We have introduced Killing-Yano tensors  in the momentum space in
   order to relate
the geometrical objects $f$ and $\tilde f$ with the
  dynamics.
   On the phase space constructed in terms of Killing-Yano tensors
  $f$ and ${\tilde f}$ we have new
  supersymmetries in the case of geodesic motion of a spinning
  particle.
 On the other hand in the case of the $n$-dimensional flat space all
Killing-Yano tensors are Nambu tensors.
  In this case we have constructed a symplectic
 structure to the space of  Killing-Yano tensors and
 the geometrical signification of these tensors was clarified.
  We found that on an even-dimensional  Riemannian  manifold
  every non-degenerate covariant constant Killing-Yano tensor
  of order two is a symplectic structure .
  We showed that it is possible to relate the
toroid dipole and quadrupole tensors to $SO$(4,2) and $Rot$(3)
generators acting in the  phase-space. This pattern is followed
also by the toroid and magnetic tensors with higher multipolarity.
Similar multipolar tensors occur in the theory of continuous media
\cite{Mis67,DM93} and  can be related to dynamical symmetry
generators in the full phase-space using a geometrical
representation valid in flat and curved spaces as we showed above.
 In this way we associate a geometrical meaning to such physical
observables of the continua.

Finding all $K{\ddot a}ller$ manifolds which have Killing-Yano
tensors is an interesting problem and it requires further
investigations \cite{nou}.

\section{Acknowledgements}
One of the authors (D.B.) would like to thank TUBITAK and NATO for
financial support and METU for the hospitality during his working
stage at  Department of Physics.


\begin{thebibliography}{99}

\bibitem{kerr} R.Penrose and M.Walker, {\em Comm.Math.Phys.} {\bf 18}, 265 
(1970). 
\bibitem{yano} K.Yano, {\em Ann.of Math.} {\bf 55}, 328 (1952).
\bibitem{hol} R.H. Rietdjik, J.W. van Holten ,
{\em Nucl. Phys.} {\bf B472}, 472 (1996).
\bibitem{dani} D.B\u aleanu and S. Codoban, {\em Gen.Rel.Grav.}
{\bf 31,no.4}, 497 (1999).
\bibitem{Gi} G.Gibbons,R.H. Rietdijk and J.W.van Holten, {\em
Nucl.Phys.} {\bf B404}, 42 (1993).
\bibitem{Lax} K.Rosquist. In {\em The Seventh Marcel Grossmann Meeting on
Recent Developmnets in Theoretical and Experimental General
Relativity, Gravitation and Relativistic Field Theories},edited
by.R.T.Jantzen and G.M.Keiser (World Scientific,
Singapore), vol.1, p.379 (1997).
\bibitem{bal}D. B\u aleanu , to be published in  {\em Il Nuovo Cimento B} 
(1999). 
\bibitem{DC74} V.M.Dubovik  and A.Cheshkov, {\em
Phys.Elem.Part\& Nucl.} {\bf 5}, 791 (1974).
\bibitem{Chan68} S.Chandrasekhar S., {\em Ellipsoidal figures of
equilibrium},  (Yale Univ, New Haven, 1969).
\bibitem{Ui70} H.Ui, {\em Prog.Theor.Phys.} {\bf 44}, 153 (1970).
\bibitem{cus68} R.Y.Cusson, {\em Nucl.Phys.} {\bf A114}, 289 (1968).
\bibitem{nambu} Y.Nambu, {\em Phys.Rev.D.} {\bf 7}, 2405 (1973).
\bibitem{takh} L.A.Takhtajan, {\em Comm.Math.Phys.}, {\bf 160}
295 (1984). L.A.Takhtajan and R.Chatterjee,  {\em
Lett.Math.Phys.} {\bf 36}, 117 (1996).
\bibitem{kram} D.Kramer {\em et al.}, {\em Exact Solutions of Einstein's
Field Equations}, (Cambridge University Press, Cambridge, 1980).
\bibitem{katayama} N.Katayama, {\em Il Nuovo Cimento B} {\bf 105 B}, 113
(1985).
\bibitem{darb}G.Darboux, {\em Bull.Sci.Math.} {\bf 6}, 14,49 (1882).
\bibitem{bali} D.B\u aleanu, {\em Helv.Acta.Phys.}  {\bf 71},
343 (1998).
\bibitem{DTOS83} V.M.Dubovik  and L.A.Tosunyan, {\em 
Phys.El.Part.\& Nuclei} {\bf 14}, 1193 (1983).
\bibitem{DT90} V.M.Dubovik  and V.V.Tugushev, {\em Phys.Rep.} {\bf 
187}, 145 (1990).
\bibitem{Mur53} Y.Murai, {\em Prog.Theor.Phys.} {\bf 9}, 147 (1953).
\bibitem{BR74} A.O. Barut and R.Raczka, {\em Theory of group
representations and their applications}, (PWN, Warshaw, 1974).
\bibitem{Fush83} V.I. Fushchich  and A.G. Nikitin, {\em The 
symmetry of Maxwell Equations}, (Naukova Dumka, Kiev, 1983).
\bibitem{Mis67} M.Mi\c sicu, {\em Mechanics of deformable continua},
( Editura Academiei, Bucharest, 1967).
\bibitem{DM93} V.M.Dubovik  and E. N. Magar , {\em J.Moscow 
Phys.Soc.} {\bf 3}, 245 (1993).
\bibitem{nou} D. B\u aleanu, in preparation.
\end{thebibliography}
\end{document}